# Lattice Disorder Effect on Magnetic Ordering of Iron Arsenides

A. S. Sefat,[1] X. P. Wang,[2] Y. Liu,[2] Q. Zou,[3] M. M. Fu,[3] Z. Gai,[3]
G. Kalaiselvan,[4] Y. Vohra,[4] L. Li,[1] D. S. Parker[1]

[1] *Materials Science & Technology Division, Oak Ridge National Laboratory, Oak Ridge, TN 37831*
[2] *Neutron Scattering Division, Oak Ridge National Laboratory, Oak Ridge, TN 37831*
[3] *Center for Nanophase Materials Sciences, Oak Ridge National Laboratory, Oak Ridge, TN 37831, USA*
[4] *Department of Physics, University of Alabama at Birmingham, Birmingham, AL 35294*

This study investigates the changes of magnetic ordering temperature via nano- and mesoscale structural features in an iron arsenide. Although magnetic ground states in quantum materials can be theoretically predicted from known crystal structures and chemical compositions, the ordering temperature is harder to pinpoint due to such local lattice variations. In this work we find surprisingly that a locally disordered material can exhibit a significantly *larger* Néel temperature ($T_N$) than an ordered material of precisely the same chemical stoichiometry. Here, a $EuFe_2As_2$ crystal, which is a '122' parent of iron arsenide superconductors, is found through synthesis to have ordering below $T_N = 195$ K (for the disordered crystal) or $T_N = 175$ K (for the ordered crystal). In the higher $T_N$ crystals, there are shorter planar Fe-Fe bonds [2.7692(2) Å vs. 2.7745(3) Å], a randomized in-plane defect structure, and diffuse scattering along the [00L] crystallographic direction that manifests as a rather broad specific heat peak. For the lower $T_N$ crystals, the $a$-lattice parameter is larger and the in-plane microscopic structure shows defect ordering along the antiphase boundaries, giving a larger $T_N$ and a higher superconducting temperature ($T_c$) upon the application of pressure. First-principles calculations find a strong interaction between $c$-axis strain and interlayer magnetic coupling, but little impact of planar strain on the magnetic order. Neutron single-crystal diffraction shows that the low-temperature magnetic phase transition due to localized Eu moments is not lattice or disorder sensitive, unlike the higher-temperature Fe sublattice ordering. This study demonstrates a higher magnetic ordering point arising from local disorder in 122.

# INTRODUCTION

A bulk magnetic transition in a quantum material is thermally-driven by spin interactions dictated by nano- and mesoscale structures, such as lattice composition, crystal structure, disorder and defects, lattice strain, chemical impurities and dopants. The parents of the iron-based superconductors with $A$Fe$_2$As$_2$ ($A$ = Ba, Sr, Ca, Eu) formula, known as '122', are semi-metallic with all five Fe 3$d$-bands crossing the Fermi level [1]. These materials have a tetragonal structure at room temperature ($a$=$b$≠$c$), and at a Néel antiferromagnetic transition temperature ($T_N$) there is a small tetragonal-to-orthorhombic structural distortion ($T_s$) where the unit cell rotates by ~ 45° within the $ab$-plane [2,3]. Below $T_N$ there is a sinusoidal modulation of Fe moments in the form of a spin-density wave (SDW) [3] described by a wave vector $q$=(½ ½ 1) in the tetragonal structure, matching the nesting vector between the electron and hole pockets at the Fermi surface [4]. The Fe spin lattice arrangement is therefore a 'stripe' (spins are antiparallel along $a$- and $c$-axes, and parallel along $b$-axis) [5-8].

It has been found that strain can induce a nematic phase transition in 122s [9] that is seen as an electronic in-plane anisotropy [10], driven by orbital (unequal occupation of $d_{xz}$ and $d_{yz}$) or spin directional order (not long range) that causes breaking of the in-plane C4 rotational symmetry and splits the $T_N$ and $T_s$ transitions. Moreover, higher $T_N$ in BaFe$_2$As$_2$ is linked to a more homogenous electronic structure [11-14], and there are local structure variations [3]. Thermally annealing of BaFe$_2$As$_2$ is found to shift and sharpen the heat capacity anomaly [15]. With disorder via electron irradiation, in BaFe$_2$(As$_{1-x}$P$_x$)$_2$ the magneto-structural transition is suppressed and the superconducting dome tracks the shift of the antiferromagnetic phase [16], while in Ba$_{1-x}$K$_x$Fe$_2$As$_2$ the antiferromagnetic and superconducting transition temperatures decrease [17]. For SrFe$_2$As$_2$, in addition to a large variability of $T_N$ values (195 to 220 K) [13,18,19], superconducting signature (filamentary $T_c$ = 21 K) can be found in strained crystals [20]. Furthermore, CaFe$_2$As$_2$ can be synthesized as entirely non-magnetic, or an antiferromagnet with a large $T_N$, achieved by staggered alleviation of local Fe−As bonds with thermal annealing [13,21,22]. For this study, we hypothesize that the local lattice details, including disorder, of EuFe$_2$As$_2$ may affect its $T_N$ value. EuFe$_2$As$_2$ is unique among the 122s for having both SDW order of Fe and Eu local moment order below 20 K with $q$=(0 0 L) [23-26]. The Eu local moments are weakly coupled to the Fe sublattice and are strong enough for a large and indirect spin-lattice coupling that can lead to a structural detwinning by a small magnetic field [27] and can cause re-entrant non-bulk superconductivity [28,29].

We have synthesized two stoichiometric EuFe$_2$As$_2$ crystals with onset of Fe ordering at $T_N$ = 195 K or $T_N$ = 175 K, and use X-ray and neutron diffraction, microscopy and spectroscopy techniques, and theory, to understand the reasons that give these ordering temperatures. Surprisingly, the higher $T_N$ crystal has a broader (usually associated with strain and disorder) specific heat peak, while the lower the $T_N$ crystal has a sharp peak. How does $T_N$ relate to lattice and topological features, and does $T_N$ value correlate with averaged lattice parameters? Would higher $T_N$ mean more homogeneous electronic structure? What are the pressure results for diminishing antiferromagnetism and potentially deriving superconductivity in each of these crystals? Our results show that although $T_{N(Fe)}$ and $T_c$ values are greatly sensitive to the lattice details and local arrangements of defect structures, Eu ordering is unaffected at $T_{N(Eu)}$≈21 K. This study demonstrates the apparently contradictory result that certain types of disorder can significantly increase magnetic order, even in bulk stoichiometric quantum material.

**EXPERIMENTAL PROCEDURE AND RESULTS**

Single crystals of EuFe$_2$As$_2$ were grown out of a mixture of Eu and FeAs excess used as liquid flux [12,30,31]. Each of these mixtures was warmed to 1180°C, then cooled (2°C/h) followed by a decanting of the FeAs excess flux at 1090°C. Two different reaction loading ratios were used to obtain EuFe$_2$As$_2$ single crystals with different $T_N$. A loading ratio of Eu:FeAs= 1:5 gives an onset transition temperature of $T_N$ = 195 K (referred to as 'crystal **a**'), while Eu:FeAs= 1:4 gives crystals with $T_N$ = 175 K ('crystal **b**'). The chemical composition of these crystals was measured with a Hitachi S3400 scanning electron microscope operating at 20 kV. Three spots (each ~80 μm area) were checked and averaged on each crystal; energy-dispersive X-ray spectroscopy (EDS) and site-refinement of single-crystal X-ray diffraction analyses indicate that both crystals are stoichiometric and are EuFe$_2$As$_2$. The phase purity of the crystals was checked by collecting data on an X'Pert PRO MPD X-ray powder diffractometer; structures were solely identified as tetragonal ThCr$_2$Si$_2$ structure type (*I4/mmm*, $Z = 2$).

Specific heat data were collected on EuFe$_2$As$_2$ single crystals, using a Quantum Design Physical Property Measurement System (PPMS); the C(T) results are shown in **Fig. 1a**. Each EuFe$_2$As$_2$ crystal exhibits two transitions: an ordering of the Fe lattice at higher temperature, followed by lower temperature ordering (≈21 K) due to Eu moments. The EuFe$_2$As$_2$ 'crystal a' exhibits the higher onset ordering temperature of $T_N$ = 195 K with a broader peak, compared to 'crystal b' giving a sharp lambda transition below $T_N$ = 175 K. The specific heat result of 'crystal a' looks similar to that reported in a pressed polycrystalline sample (with respect to peak width, height, and transition temperature) [23].

To analyze crystal structures, single crystal X-ray diffraction data were collected at room temperature on a Rigaku XtaLAB PRO diffractometer (Mo Source, K$_\alpha$ =0.71073 Å) equipped with a DECTRIS Pilatus 200K area detector. Data collection and reduction used the CrysAlisPro program [32]. Crystal structure refinements were performed using the SHELX-2014 program [33]. **Fig. 1b** (insets) shows the pictures of crystals for the data collection, stuck on top of needles. Crystals 'a' and 'b' appeared similar visually, with sheet morphologies and dimensions of ~0.15 × 0.15 × 0.02 mm$^3$ or smaller in *a*, *b*, and *c* crystallographic directions, respectively; [001] direction is perpendicular to the plane of the plate. **Fig. 1b** gives the precession images of [HK0] and [H0L] reciprocal lattice planes, showing lattice strain effects associated with smeared diffraction spots in the L direction for 'crystal a'; this crystal has a relatively broad C(T) peak ($T_N$ = 195 K) from the Fe ordering. A broader peak of specific heat is usually associated with lattice strain and disorder. **Table 1** shows the lattice parameters and refinement results of the X-ray structure measured at room temperature. The EuFe$_2$As$_2$ sample with higher $T_N$ ('crystal a') gives a lower overall *a*-lattice parameter and a shorter planar Fe−Fe distance of 2.7692(2) Å, indicating a contraction of 0.0053(4) Å for the in-plane Fe−Fe bond when compared to that of 2.7745(3) Å for the lower $T_N$ ('crystal b').

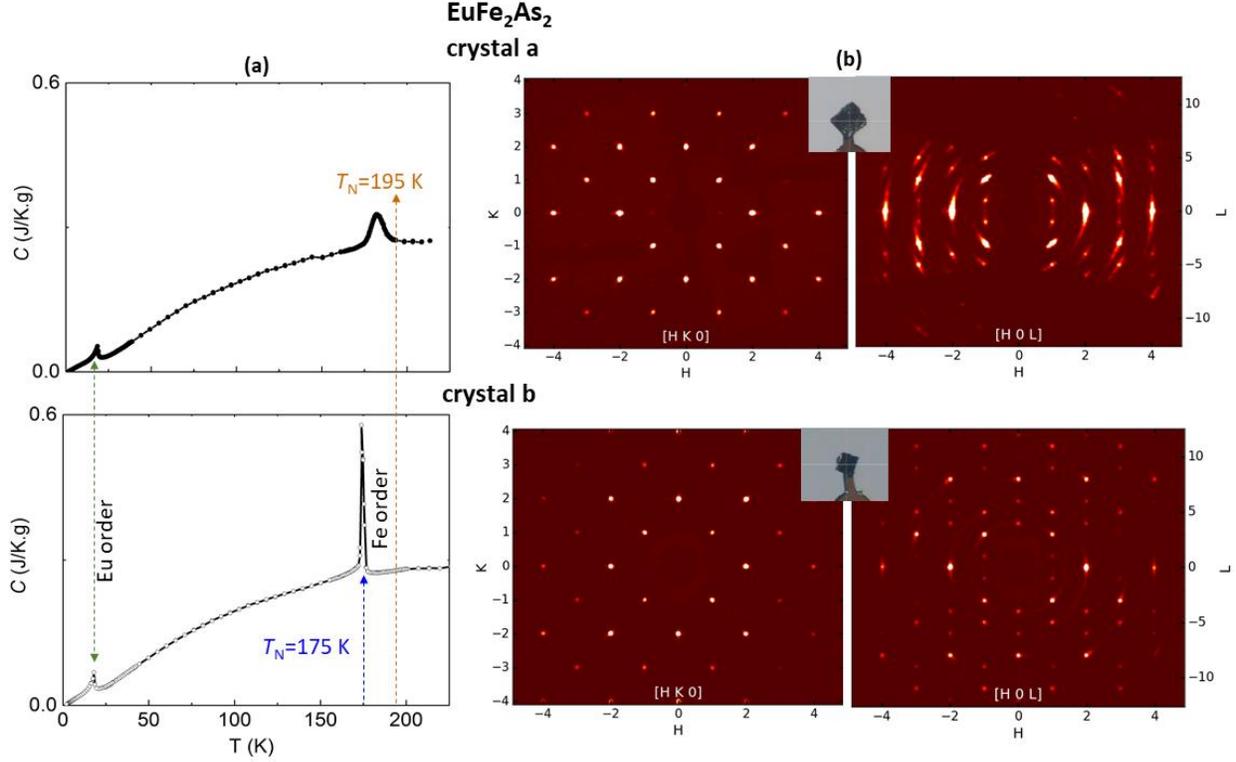

**Fig. 1**: Measurement of EuFe$_2$As$_2$ 'crystal a' and 'crystal b'. (a) Specific heat results with transitions associated with Eu ($T_N \approx 21$ K) and Fe ordering ($T_N = 195$ K or 175 K), C(T). (b) Precession images showing the [HK0] and [H0L] reciprocal lattice planes reconstructed from single crystal X-ray diffraction patterns measured at room temperature. Smearing in diffraction spots are more prevalent and are noted for image [H0L] in 'crystal a'.

**Table 1**: Single-crystal X-ray diffraction refinement on the two crystals of EuFe$_2$As$_2$: 'crystal a' ($T_N$ = 195 K), 'crystal b' ($T_N$ = 175 K).

| EuFe$_2$As$_2$ sample ID | 'crystal a' | 'crystal b' |
|---|---|---|
| $a$ (Å) | 3.9162(3) | 3.9238(4) |
| $c$ (Å) | 12.104(3) | 12.105(2) |
| As at (0, 0, $z$); 4e | | |
| $z$ | 0.36256(13) | 0.36255(11) |
| $U_{iso}$ | 0.0104(6) | 0.0109(6) |
| arsenic height, Å | 1.3624(10) | 1.3624(8) |
| site occupancy | 0.96(9) | 0.97(8) |
| Fe at (½, 0, ¼); 4d | | |
| $U_{iso}$ | 0.0106(6) | 0.0107(6) |
| Fe–Fe distance, Å | 2.7692(2) | 2.7745(3) |
| site occupancy | 0.97(9) | 0.97(8) |

Neutron diffraction measurements were carried out using the TOPAZ single crystal diffractometer at the ORNL Spallation Neutron Source (SNS), which uses the wavelength-resolved Laue technique with an extensive array of neutron time-of-flight area detectors for efficient 3-dimentional $Q$ space mapping of Bragg and diffuse scattering patterns originating from magnetic and nuclear phase transitions [34]. $EuFe_2As_2$ 'crystal a' and 'crystal b' had dimensions of 3.25×2.50×0.20 $mm^3$ and 2.75×2.63×0.30 $mm^3$, respectively, and were attached on a MiTegen loop using super glue for data collection at room temperature and 95 K. Each of the samples was oriented with high precision for volumetric sampling of Bragg peaks in specific directions with the CrystalPlan program [35]. At room temperature, the (0 2 0)$_{Tetragonal}$ peak profiles for the two $EuFe_2As_2$ crystals are shown in **Fig. 2**, demonstrating the peak false-color maps and their corresponding peak intensity profiles along the crystallographic $c$-direction. The peak from 'crystal a' is broad and shows extensive diffuse lines along the L direction, which is consistent with that observed in X-ray diffraction, and gives broader and higher $T_N$ in C(T).

As shown in **Fig. 1b** and **Fig. 2**, the streaks for the (2 0 0) peak along L direction for 'crystal a' is much more pronounced than that of 'crystal b', which could be induced by strain, stacking fault or disorder in the bulk single crystal sample. Since the refined average $c$ lattice parameter remains essentially unchanged for both crystals (**Table 1**), the much more pronounced streaks for 'crystal a' are likely from microstrain caused by randomly distributed defects in 'crystal a', which is evident from scanning tunneling microscopy/spectroscopy (STM/S) measurement (see **Fig. 3** below). The split of the peaks along $c$ also indicates twinning and multidomain structures of 'crystal a'. In particular, we note from **Fig. 2** that the full-width at half maximum (FWHM) for the (2 0 0) peak along L direction is some 0.13 – 0.27 reciprocal lattice units larger than that of crystal 'b'. If one assumes that in the ordered crystal crystal 'b' the finite (i.e. non-zero) FWHM is largely due to measurement resolution limits, the FWHM for crystal 'a' would correspond to a ~2% – 5% local variation in the lattice constant $c$.

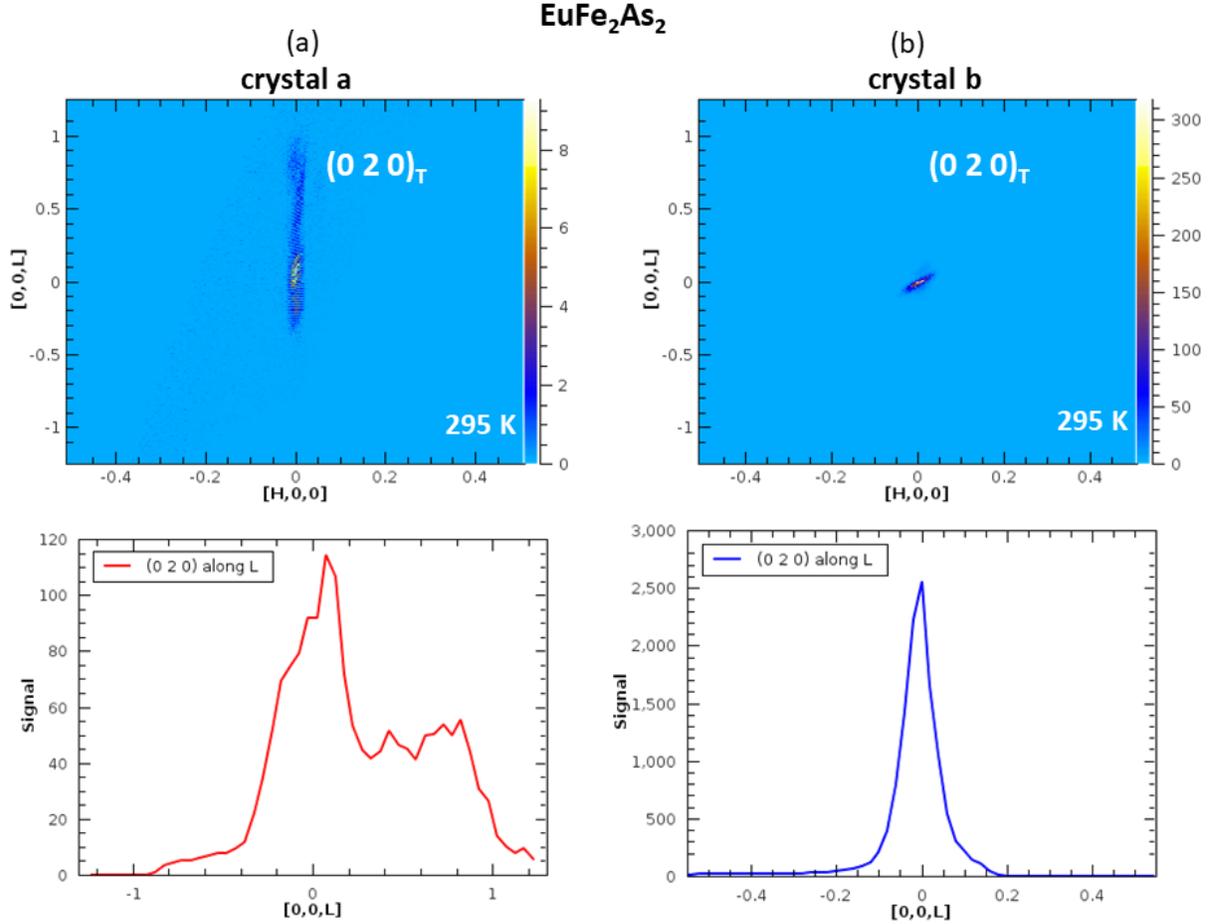

**Fig. 2**: Tetragonal (0 2 0) peak profiles for the two EuFe$_2$As$_2$ crystals labeled 'a' and 'b' at room temperature, showing the peak false color maps, and their corresponding peak intensity profiles along the crystallographic *c* direction. Note the difference in scales on both axes. The peak from 'crystal a' is substantially broadened showing extended diffuse lines.

For these crystals, neutron diffraction results were additionally carried out at 95 K, to confirm the Fe magnetic peak below $T_N$ in accordance with the propagation vector $q$=(1 0 1)$_{Orthorhombic}$=(½ ½ 1)$_{Tetragonal}$ and the stripe arrangement of the Fe spin lattice [5-8] (see Figure **S1a** in **S**upplementary). Additionally, variable temperature scans of the Eu magnetic peak at (0 3 0) for both crystals were performed below 30 K using the elastic diffuse scattering spectrometer CORELLI at SNS [33] and showed the same temperature dependence of Eu magnetic peak as (0 3 0) with $T_N \approx 21$ K (see Figure **S2**). In contrast, the Néel transition temperature for Fe is clearly lattice and disorder sensitive. The high-temperature phase transition is due to the ordering of the transition metal Fe moments, whereas the low-temperature phase transition is due to the ordering of the localized Eu moments [23]. The divergent behavior of the Fe and Eu sublattices from 'crystal a' to 'b' – the former *sensitive* to lattice disorder, the latter *insensitive* – is of interest here. One plausible hypothesis is that the Eu 4*f* moments are much more localized, near the Eu nucleus, than the Fe moments which are more itinerant and physically spread out in more of the unit cell. This would suggest that the Eu 4*f* moments are less subject to nano-strain than the Fe 3*d* moments, and the relatively isolated location of Eu between the layers (as opposed to the tightly bound FeAs layers) would enhance this possibility. Supporting this assertion is the relative insensitivity of Eu ordering temperature to pressure (see Fig. **S4**), again unlike the Fe sublattice ordering. Indeed the Eu magnetic order survives to higher pressure than the Fe order, despite occurring at much lower temperatures.

To further investigate the local origin of the bulk $T_N$ differences between the two EuFe$_2$As$_2$ crystals, surface topography and electronic structures were investigated using STM/S on in-situ low-temperature cleaved surfaces. The two sets of EuFe$_2$As$_2$ 'crystal a' and 'crystal b' were cleaved in an ultra-high vacuum at ~ 78 K and then immediately transferred to the STM/S head precooled to 4.2 K without breaking vacuum. The STM/S experiments were carried out using a scanning tunneling microscope with base pressure better than $2\times10^{-10}$ Torr, with mechanically cut Pt-Ir tip. All Pt-Ir tips were conditioned on clean Au (1 1 1) and checked using the topography, surface state, and work function before each measurement. The STM/S were controlled by the SPECS Nanonis control system. Topographic images were acquired in constant current mode with bias voltage applied to samples, and tip grounded. All the spectroscopies were obtained using the lock-in technique with a modulation of 1 mV at 973 Hz on bias voltage, dI/dV. Current-imaging-tunneling-spectroscopy were collected over a grid of pixels at bias ranges around Fermi level using the same lock-in amplifier parameters. The survey on multiple large areas of both samples shows the coexistence surface reconstructions on 'crystal a' and 'crystal b', as shown in **Fig. 3**. Using a similar method as STM report on Co-doped BaFe$_2$As$_2$ [37], the 2 × 1 and √2 × √2 reconstructed surfaces can be assigned to arsenic termination and europium termination, respectively, as shown in **Fig. 3a**. While the 2 × 1 surfaces (As termination) of both crystals (**Fig. 3b**, **c**) are very similar, the √2 × √2 surfaces (Eu termination) of the two crystals (**Fig. 3d**, **e**) are very different. The atomic resolved images in the insets of **Fig. 3d** and **e** from the well-ordered √2 × √2 reconstructed areas of the two crystals are similar, but the arrangements of the defects on the surfaces are rather different. In 'crystal a', surface defects (one of the defects is marked with a black arrow in **Fig. 3d**) are essentially randomly distributed on the surface, but in 'crystal b', large amount of defects prefer to form into chains (one of the chains is marked with a white arrow in **Fig. 3e**). By analyzing atomically resolved images around these areas, we found the chains are aligned on the antiphase boundaries. Because the surface reconstruction is 2 × 1 or √2 × √2, those domains can shift by 1 to form antiphase boundaries. Although antiphase boundaries also exist on the 'crystal a' surface (one of the antiphase boundaries is marked with a green arrow in **Fig. 3d**), defects in 'crystal a' do not segregate along the boundaries. The electronic properties revealed by the STS from the surfaces are consistent with the morphological observation. In **Fig. 3f**, the average electronic local density of states (LDOS) of 2 × 1 surfaces (As termination) over large areas are almost identical for both crystals, but the LDOS from the two √2 × √2 surfaces (Eu termination) in **Fig. 3g** have significant differences. This shows that the different arrangements of the surface defects in the two crystals change the electronic structure dramatically. Given the preparation of these surfaces by cleaving the crystals in ultra-high vacuum at low temperatures, this difference is in electronic structure that is generally reflective of the effects of these defects in the bulk crystals.

For both EuFe$_2$As$_2$ crystals, high-pressure electrical-resistance measurements were performed using a diamond anvil cell (see **Fig. S3**, **S4**), to explore the differences of pressure effects on the two crystals that have different $T_N$, under the same experimental setup. Although the feature due to Eu ordering is not changed for either crystals up to ~ 4 GPa, Fe ordering is greatly sensitive to pressure and the rate of $T_N$ suppression for both crystals is similar. For 'crystal b' with smaller $T_N$ = 175 K, the drop in resistivity is noticed at lower pressure of 2.5 GPa, compared to 'crystal a,' with a drop in resistivity at 3.2 GPa. EuFe$_2$As$_2$ with sharper, but lower $T_N$, may be prone to a higher superconducting dome (**S4**).

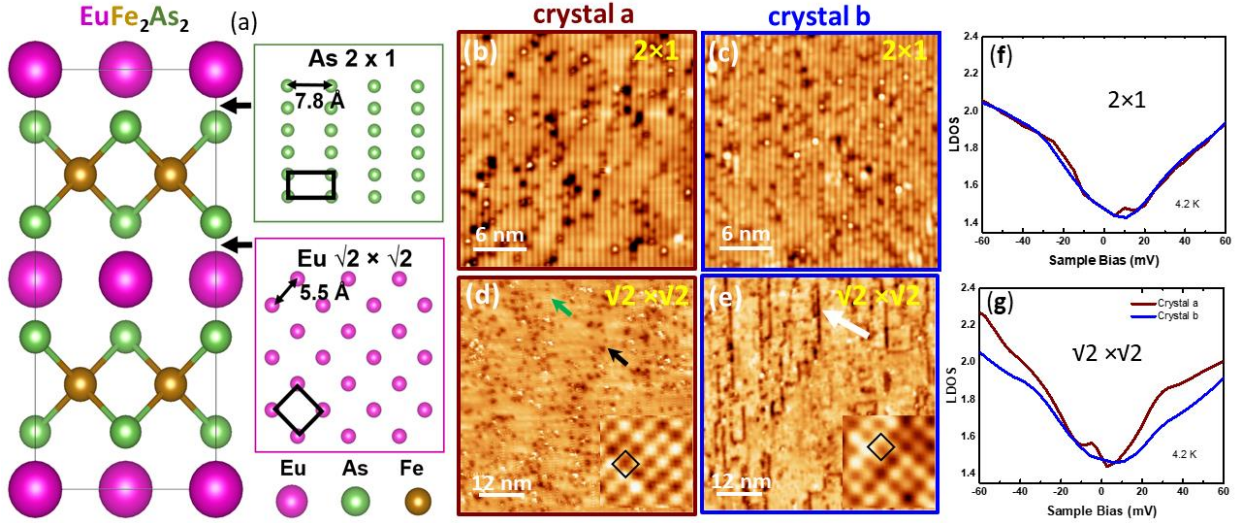

**Fig. 3**: (a) The structure model of EuFe$_2$As$_2$ and the surface reconstruction models of As 2 × 1 and Eu √2 × √2 termination. The black boxes outline the unit cells. (b, c) Topographic images of 2 × 1 surface reconstruction from 'crystals a' (-500 mV, 100 pA) and 'crystal b' (-1 V, 100 pA), respectively. (d, e) Topographic images of √2 × √2 surface reconstruction from 'crystals a' (-60 mV, 400 pA) and 'crystal b' (-20 mV, 800 pA), respectively. Insets show the atomic resolved images in 3×3 nm$^2$ size and black boxes are the unit cells of √2 × √2 surface reconstruction. (f, g) Comparison of average LDOS from 'crystals a' (red) and 'crystal b' (blue), for 2 × 1 and √2 × √2 surface reconstruction at 4.2 K.

**FIRST PRINCIPLES CALCULATIONS**

To understand the sample-to-sample variation in Néel temperatures, we have conducted first principles theory calculations of the effect of nano-scale strain on the magnetic order, specifically on the strength of interlayer coupling. First principles calculations were performed using the linearized augmented plane wave (LAPW) density functional theory code WIEN2K [38]. As in our previous work [14] we have used the local density approximation (LDA), with arsenic height and lattice parameters taken directly from our experimental XRD refinements (**Table 1**), except as stated below. An RK$_{max}$ value of 8.0 – the product of the smallest sphere radius and largest planewave expansion wavevector – was used. To avoid the generally confounding complexities of 4$f$ physics (here we are focused on the Fe magnetic behavior) we have followed our previous theoretical work [39] and performed the all-electron calculations with the isoelectronic substitution of Sr for the divalent Eu, but retained our lattice parameters and internal coordinates for EuFe$_2$As$_2$. The effect of these choices is to allow a direct evaluation of the Fe sublattice magnetic properties without the need to apply a correlated (i.e. LDA+U) approach to deal with the Eu 4$f$ electrons, which play little or no role in the Fe magnetism. Sphere radii of 2.5 Bohr were used for Sr, and between 2.26 and 2.30 for Fe (depending on volume), and between 2.15 and 2.22 for As. Identical sphere radii were used for the calculations at given volume, relevant for assessing the interlayer coupling.

Neutron scan of 'crystal a' along the [00L] direction shows clear evidence of disorder, with a rather broadened peak around the (0 0 0) point and a substantial 'shoulder' feature extending nearly to the [001] point (**Fig. 2a**). While there are many possible sources of such disorder, ranging from vacancies and anti-site defects to small grain sizes, here we explore the idea that $c$-axis strain in 'crystal a' is ultimately responsible for the rather broad neutron diffraction peaks. We have therefore conducted calculations of the interlayer magnetic coupling. Here we report the energy difference, per unit cell, between the ground-state stripe magnetic order (which is antiferromagnetically coupled along the $c$-axis) and a state with ferromagnetic $c$-axis coupling. We have done this at the experimental lattice parameter of 12.104 Å, and in addition for $c$ values 5% larger and smaller than this value. For the experimental $c$-lattice parameter, we find an interlayer coupling energy difference of 7.1 mRyd/u.c., which increases slightly to 7.7 mRyd for the +5% $c$ value, but decreases sharply to some 4.5 mRyd for the -5% $c$ value. These values are in qualitative agreement with the previous work on $BaFe_2As_2$ [40], where it was found that the application of pressure decreases the interlayer coupling. We mention also that the calculated magnetic moments decrease substantially for the smaller cell, and correspondingly increase for the larger cell. In the antiferromagnetically coupled layer state, for the smallest cell these moments are 1.25 $\mu_B$ per Fe, for the experimental cell the value is 1.63 $\mu_B$, and for the largest cell is 1.92 $\mu_B$. Neutron diffraction results give Fe moment of ~1 $\mu_B$ [8].

The point of these results is that interlayer strain can plausibly affect the magnetism in a substantial manner, consistent with the disordered lattice evident in diffraction and the observed onset temperature. What is more difficult to understand is the apparent onset increase in $T_N$ with disorder – i.e., 'crystal a' compared with 'b'. This is particularly true given that the strained calculations show larger *decreases* in magnetic order (measured as the interlayer coupling) with *compressive* strain than increases with tensile strain. Note also that our calculations of the magnetic ordering energy, defined as the difference in energy between the stripe ground state and 'checkerboard' excited state, find no significant difference (< 1 %) resulting from the small *planar* lattice parameter change from 'crystal a' to 'b' (the experimental structures were used here). In the simplest approximation based on this T=0 K approach, these results would suggest that this planar lattice change is not the source of the $T_N$ variation. However, given that applied pressure (thereby reducing lattice parameters) is well-known as an effective means of Néel temperature suppression, it remains possible that this effective strain is relevant here. Arguing against this, however, is the fact that the *smaller* lattice parameter 'crystal a' has the *larger* Néel point, whereas the application of pressure, thereby yielding smaller lattice constants, generally *lowers* the Néel point.

Since the obvious explanations of the $T_N$ increase, based on disorder-induced structure modification, do not explain this rather unusual behavior, it is necessary to consider other possibilities. For example, in materials near a magnetic instability, it has recently been posited [41] that charge doping can induce a magnetic transition based on Stoner physics, and one could plausibly imagine a similar enhancement in magnetic character of an already magnetic material. However, the generally stoichiometric character of both samples studied here argues against such a possibility. Instead, we suggest the following: spin fluctuations are known to be exceptionally strong in the iron arsenides, and in fact are a leading candidate for the interaction causing superconductivity. It is commonly believed [42] that these fluctuations also play an important role in reducing the Néel point of these materials from a much higher 'bare' value. It is also known from several recent theory works [43,44] that disorder can play a substantial role in weakening spin fluctuations and inducing magnetism, and this role is especially prominent [45] in the case of *stripe* magnetic order, as is observed in the iron-arsenides. Putting such arguments together, we suggest that the disorder evident in our specific heat and neutron data is weakening the previously very strong spin fluctuations and thereby enhancing the magnetic order.

One means of testing this theory would be to employ the quasi-particle interference (QPI) techniques recently applied to LiFeAs [46]. While in that work it was the superconductivity (and its coupling to a bosonic excitation) that was studied, we anticipate that a measure suitably adapted to the case of the antiferromagnetic order here could also yield insight. In particular, if reduced spin fluctuations are in fact at work here, the momentum and energy dependent differential conductance g($\mathbf{k}$,ω) should show signatures of reduced scattering in the higher $T_N$. In particular, we would generally expect sharper (as a function of momentum and energy) peaks in this conductance in this material, when measured at the same temperature as the lower Néel point crystal. Future experimental studies would be highly beneficial are to assess this possibility.

**CONCLUSION**

We find that disorder-related lattice variability drives significantly different Fe Néel ordering temperatures in the stoichiometric EuFe$_2$As$_2$ crystal, with the highly unusual result that the *more* disordered crystal exhibits the *higher* Néel point. The diffuse scattering that is seen along the [00L] direction in X-ray and neutron diffraction measurements for the higher $T_N$ crystal may well be the origin of the broadened specific heat peak and the suppressed superconducting dome. The surface topography and electronic structure studies show that there is a clear difference in electronic and defect structures between the two $T_N$ crystals, with defect states dominating and elevating the LDOS for the higher $T_N$ 'crystal a'. Although the two crystals have a similar number of defects, their segregation around the antiphase boundaries in 'crystal b' largely decreases the number of individual defects on the surface, which may produce a higher-$T_c$ superconductor with pressure. This study thereby demonstrates the modification of $T_N$ in an iron arsenide by controlled disorder, thus explaining the observation of substantial ordering temperature variations for these stoichiometric quantum materials in the literature.

This manuscript has been authored by UT-Battelle, LLC under Contract No. DE-AC05-00OR22725 with the U.S. Department of Energy. The United States Government retains and the publisher, by accepting the article for publication, acknowledges that the United States Government retains a non-exclusive, paid-up, irrevocable, world-wide license to publish or reproduce the published form of this manuscript, or allow others to do so, for United States Government purposes. The Department of Energy will provide public access to these results of federally sponsored research in accordance with the DOE Public Access Plan.

The research is primarily supported by the U.S. Department of Energy (DOE), Office of Science, Basic Energy Sciences (BES), Materials Science and Engineering Division. Neutron scattering experiments at Oak Ridge National Laboratory (ORNL) were supported by the Scientific User Facilities Division, Office of BES, DOE.

# Supplementary Material

for

# Lattice Disorder Effect on Magnetic Ordering of Iron Arsenides

A. S. Sefat,[1] X. P. Wang,[2] Y. Liu,[2] Q. Zou,[3] M. M. Fu,[3] Z. Gai,[3]
G. Kalaiselvan,[4] Y. Vohra,[4] L. Li,[1] D. S. Parker[1]

[1] *Materials Science & Technology Division, Oak Ridge National Laboratory, Oak Ridge, TN 37831*
[2] *Neutron Scattering Division, Oak Ridge National Laboratory, Oak Ridge, TN 37831*
[3] *Center for Nanophase Materials Sciences, Oak Ridge National Laboratory, Oak Ridge, TN 37831, USA*
[4] *Department of Physics, University of Alabama at Birmingham, Birmingham, AL 35294*

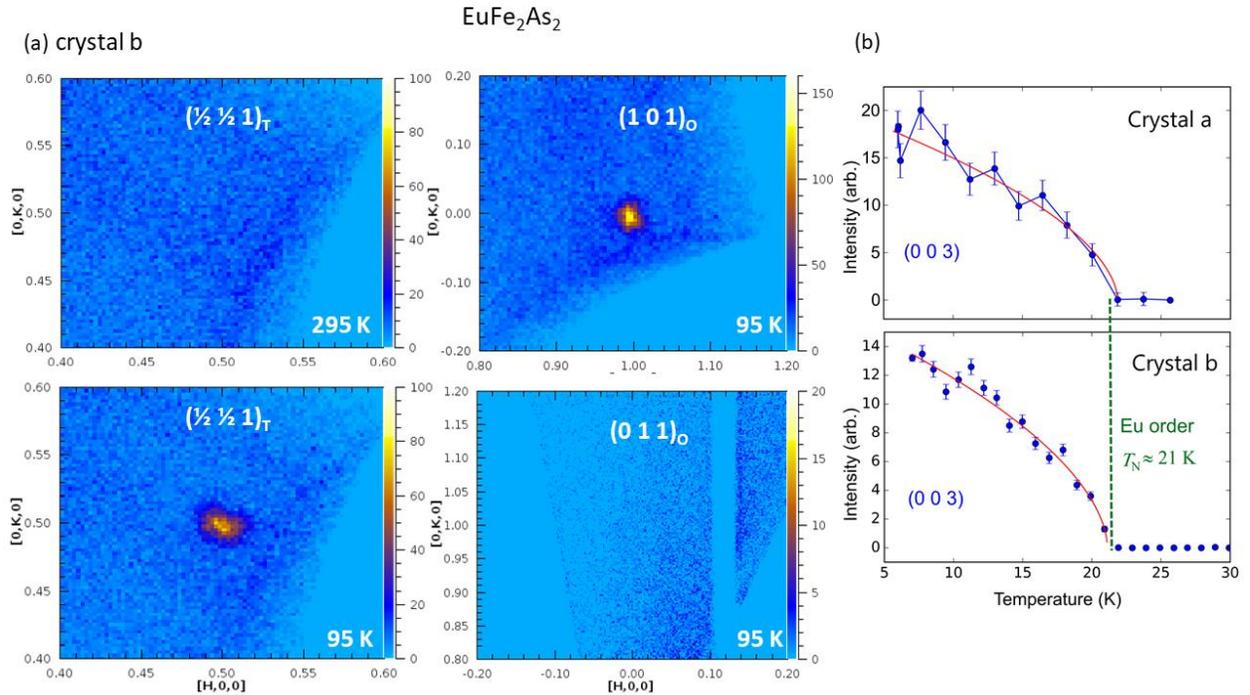

**Fig. S1**: For $EuFe_2As_2$: (a) Neutron single crystal diffraction showing the appearance of Fe magnetic peak for 'crystal b' below $T_N$ in accordance with the propagation vector $(1\ 0\ 1)_O = (½\ ½\ 1)_T$. Left: There is no Fe magnetic peak in the tetragonal phase at room temperature, while Fe magnetic superlattice peak is seen at 95 K indexed in $(½\ ½\ 1)_T$. Right: The same Fe magnetic peak indexed as $(1\ 0\ 1)_O$ in the low temperature orthorhombic phase after basis transformation (***a***+***b***, -***b***+***a***, ***c***), while absence of $(0\ 1\ 1)_O$ is consistent with 'stripe' arrangement of Fe spin lattice known for $EuFe_2As_2$ [1]. The Fe spins are antiparallel along *a*- and *c*-axes, and parallel along *b*-axis. No twinned domains are observed for 'crystal b'. (b) Temperature dependence of the Eu $(0\ 0\ 3)$ magnetic reflection for crystals 'a' and 'b'. Both samples exhibit approximately the same ordering behavior in terms of ordering temperature (≈21 K) and critical exponents (0.28).

For EuFe$_2$As$_2$ crystals, neutron diffraction results were carried out at 95 K using TOPAZ single-crystal diffractometer, with the temperature controlled by an Oxford CryoStream with $l$-N$_2$ flow. The appearance of an Fe magnetic peak below $T_N$ is confirmed in accordance with the propagation vector $q$=(1 0 1)$_O$=(½ ½ 1)$_T$. Additionally, the absence of (0 1 1)$_O$ reflection at temperatures below $T_N$ are consistent with the assignment of Fe spins are antiparallel along $a$- and $c$-axes, and parallel along $b$-axis of the orthorhombic cell; these results are shown for 'crystal b' in **Fig. S1(a)**. The same EuFe$_2$As$_2$ single crystal samples 'a' and 'b' used for the TOPAZ experiments were remounted using an aluminum pin on the CORELLI cryostat for data collection below 30 K. Peak intensities for individual temperature scans used ~ 7 minutes of neutron beam time, with 0.5 Columbus of proton-charge on target as the stopping criteria for data collection at each temperature. The temperature dependent Eu magnetic peak of (0 3 0) shows approximately the same ordering for both crystals, with critical exponents 0.28(5) for 'crystal a' and 0.28(3) for 'crystal b.' These results are shown in **Fig. S1(b)**.

To better understand how the local chemical structure changes the electronic structure on √2 × √2 surfaces, the spatial distributions of LDOS at ± 60 mV from the two EuFe$_2$As$_2$ crystals are plotted in **Fig. S2** along with the simultaneously acquired topographic images. For 'crystal a', the intensities of the LDOS maps (**Fig. S2b**, **c**) are relatively higher and more uniform compared to 'crystal b' (**Fig. S2e**, **f**). What marks the difference between the two crystals are the large quantities of low density-of-state areas in 'crystal b' (green areas in **Fig. S2e** and dark blue areas in **Fig. S2f**), those are areas with fewer vacancy defects in the topographic images. It is interesting to note that the LDOS intensity and influential territory around individual vacancy-defects are much higher than that of vacancy chains. This observation explains the different average LDOS in these crystals. Although the two surfaces of the crystals have a similar number of Eu vacancies, their segregation around the antiphase boundaries in 'crystal b' largely decreases the amount of individual vacancies on the surface. The relatively large amount of random individual vacancies in 'crystal a' largely elevates the density of states.

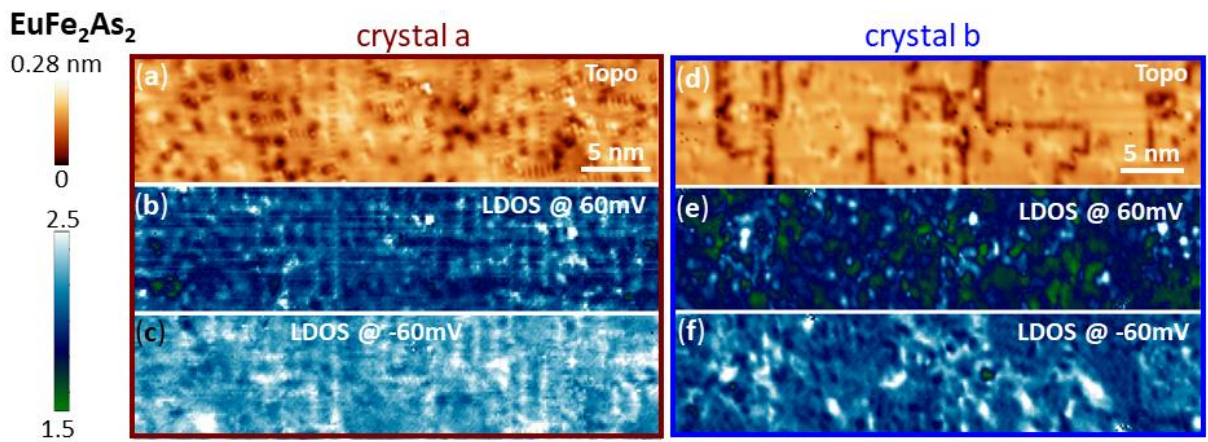

**Fig. S2**: EuFe$_2$As$_2$ local structures. (a, b, c) Topographic image, LDOS at 60 mV and -60 mV of √2 × √2 reconstructed surface on 'crystals a' (-60 mV, 800 pA). (d, e, f) Topographic image, LDOS at 60 mV and -60 mV of √2 × √2 reconstructed surface on 'crystal b' (-60 mV, 800 pA).

For both EuFe$_2$As$_2$ single crystals, high-pressure electrical-resistance measurements were performed using a diamond anvil cell (**Fig. S3a**), similar to earlier reports [2-4]. In situ, one of the anvils was symmetrically arrange and deposited with eight tungsten microprobes encapsulated in a homoepitaxial diamond film and are exposed only near the tip of the diamond to make contact with the EuFe$_2$As sample at high pressure. The designer diamond anvil was beveled with flat diameter of 100 μm and culet diameter of ~ 300 μm. The distance between two opposite leads was ~ 50 μm. The gasket was made from a 240 μm thick hardened spring steel foil pre-indented to 80 μm with a 120 μm diameter hole electrospark drilled through the center of the preindented region. Apart from eight electrical leads, two leads were used to set constant current through the sample and the two additional leads were used to monitor the voltage across the sample. A steatite pressure medium was employed in the electrical resistance measurements to assure electrical insulation of the sample from metallic gasket. The pressure was monitored by the ruby fluorescence technique and care was taken to carefully calibrate the ruby R$_1$ and R$_2$ emission in both samples at low temperatures. Although there are reports of the application of pressure on EuFe$_2$As$_2$ causing suppression of $T_N$ and superconductivity onset at pressure values of ~ 2 to 3 GPa [5,6], here we want to explore the differences of pressure effects on the two crystals with different $T_N$. A steatite pressure medium was employed in the electrical resistance measurements to assure some pressure uniformity. The temperature dependence of resistivity results is shown in **Fig. S3e**, up to ~ 20 GPa. For these crystals, clear anomalies due to Fe and Eu ordering are seen in the data. Although the feature due to Eu ordering is not changed for either crystals up to ~ 4 GPa, Fe ordering is greatly sensitive to pressure and the rate of $T_N$ suppression for both crystals is similar. For 'crystal b' with smaller $T_N$= 175 K, the drop in resistivity is noticed at lower pressure of 2.5 GPa, compared to 'crystal a,' with $T_c$ setting in at 3.2 GPa. EuFe$_2$As$_2$ with sharper but lower $T_N$ ('crystal b') gives a slightly higher superconducting dome, summarized in $T$-P phase diagram in **Fig. S4**.

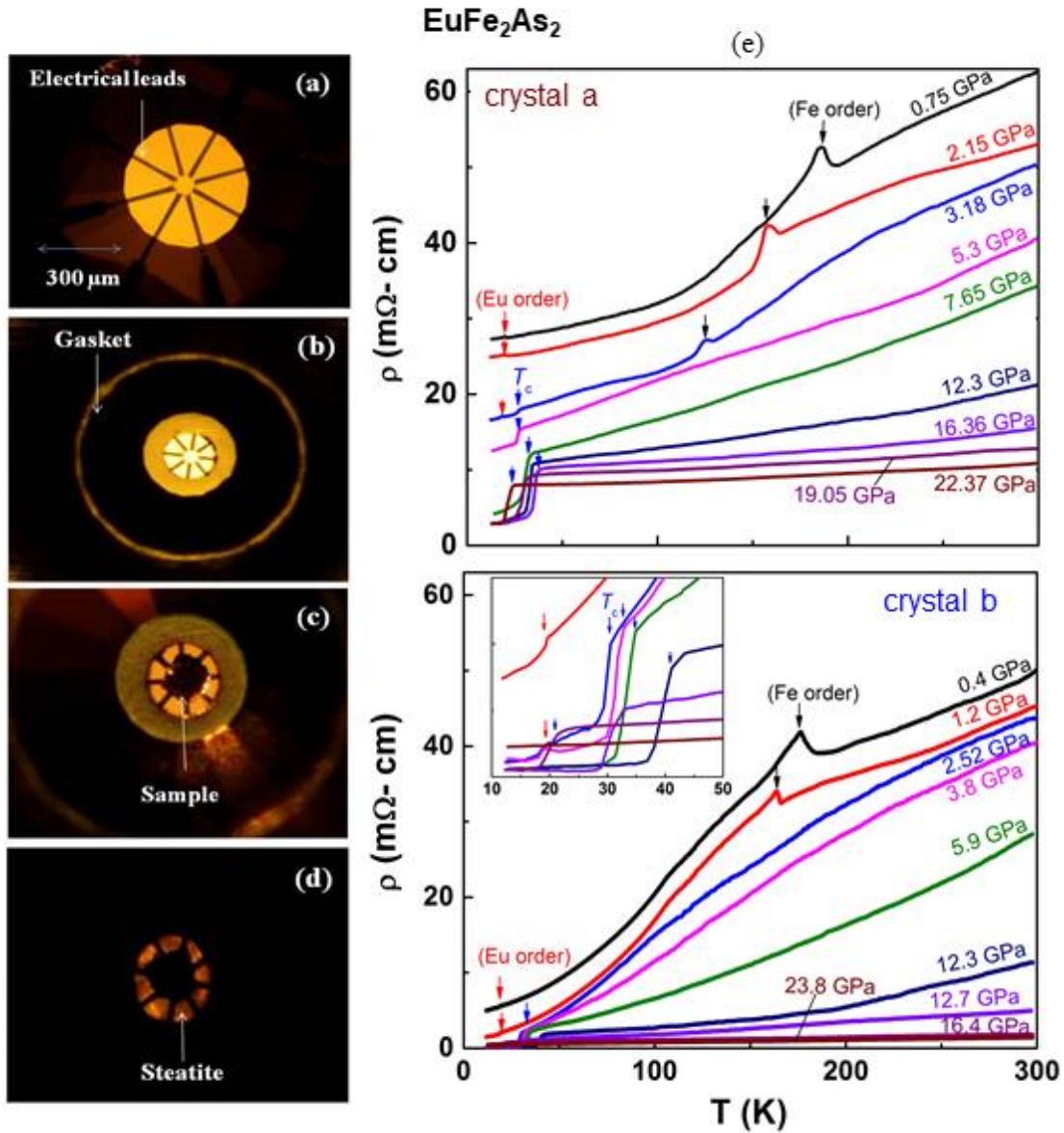

**Fig. S3.** (a) Eight probe designer diamond anvil used in high-pressure resistivity measurements of EuFe$_2$As$_2$ single crystal. The current and voltage contacts in the four probe electrical measurements are indicated. (a) The inset is the close-up of the diamond culet showing the eight shiny tungsten metal probes emerging near the center to make contact with the sample at high pressures. The metal probes are embedded in a chemical vapor deposited diamond layer elsewhere except for the contacts indicated. (b) Metallic gasket mounting in the designer diamond, (c) and (d) EuFe$_2$As$_2$ crystal loaded with a steatite pressure medium, thereby electrically insulating the sample from the gasket. (e) Temperature and pressure dependence of the electrical resistivity of EuFe$_2$As$_2$ 'crystal a' (top) and 'crystal b' (bottom).

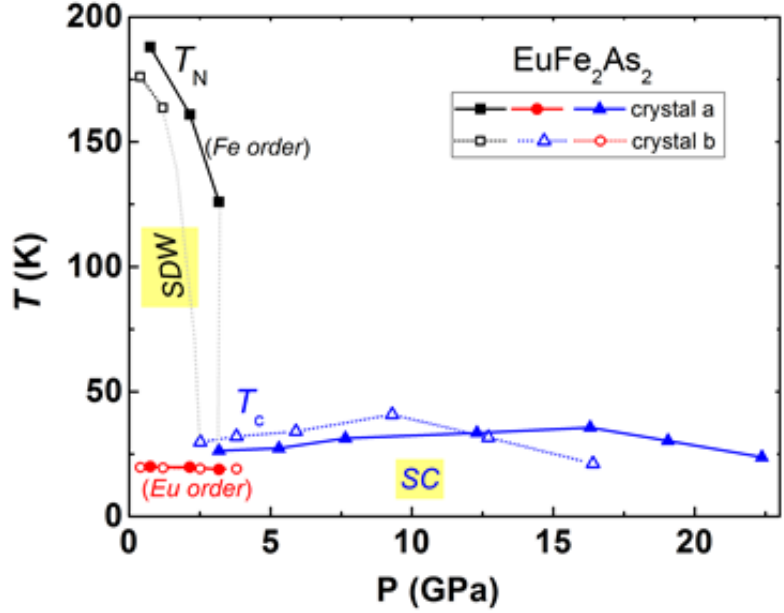

**Fig. S4**: Temperature-pressure phase diagram for the two EuFe$_2$As$_2$ crystals with different $T_N$ at ambient pressure.

18